\documentclass[12pt]{article}
\usepackage{amsmath}
\usepackage{psfrag,epsf}
\usepackage[pdftex]{graphicx}
\usepackage{enumerate}
\usepackage{natbib}
\usepackage{color}
\usepackage{comment}
\usepackage{ifthen}
\usepackage{comment}
\usepackage{ifthen,subcaption}
\usepackage{xcolor, soul}
\usepackage{tabularx, booktabs}
\usepackage{float}

\usepackage{url} 
\newcommand{\blind}{1}
\definecolor{ao(english)}{rgb}{0.0, 0.5, 0.0} 
\definecolor{awesome}{rgb}{1.0, 0.13, 0.32}
\usepackage{epstopdf}
\newcommand\numberthis{\addtocounter{equation}{1}\tag{\theequation}}
\newcommand{\compos}{\mbox{$\bot\!\!\!\bot$}}

\newtheorem{example}{Example}

\newtheorem{definition}{Definition}
\newtheorem{thm}{Theorem}
\newtheorem{cor}{Corollary}

\newtheorem{remark}{Remark}

\begin{document}

\def\spacingset#1{\renewcommand{\baselinestretch}%
{#1}\small\normalsize} \spacingset{1}


\if1\blind
{
  \title{\bf Graphical Models for Processing Missing Data}
  \author{Karthika Mohan\thanks{
    The authors gratefully acknowledge support of this work by grants from
    NSF IIS-1302448, IIS-1527490 and IIS-1704932; ONR N00014-17-1-2091;
    DARPA W911NF-16-1-0579.}\hspace{.2cm}\\
    Department of Computer Science, University of California Los Angeles\\
    and \\
    Judea Pearl \\
		Department of Computer Science, University of California Los Angeles}
  \maketitle
} \fi

\if1\blind
{
  \bigskip
  \begin{center}\texttt{}
    {\LARGE\bf Graphical Models for Processing Missing Data}
\end{center}
  \medskip
} \fi

\bigskip
\begin{abstract}
This paper reviews recent advances in missing data
research using graphical models to represent multivariate
dependencies. We first examine the limitations of traditional
frameworks from three different perspectives:
\textit{transparency, estimability and testability}.
We then show how procedures based on graphical
models can overcome these limitations and provide
meaningful performance guarantees even when data are
Missing Not At Random (MNAR). In particular, we identify
conditions that guarantee consistent estimation in broad categories of missing data problems,
and derive procedures for implementing this estimation.
Finally we derive testable implications for missing data 
models in both  MAR (Missing At Random) and MNAR categories.
\end{abstract}

\noindent%
\vfill

\newpage
\section{Introduction}
\label{sec:intro}

Missing data present a challenge in many branches
of empirical sciences.  Sensors do not always work
reliably, respondents do not fill out every question
in the questionnaire, and medical patients are often unable to recall episodes, treatments or outcomes.
The statistical literature on this problem is rich and abundant and has resulted in
 powerful software
packages such as MICE in R, Stata, SAS and SPSS which offer various ways of handling missingness.  
Most practices are based on the seminal  work of 
\cite{Rubin_missing} who formulated procedures
and conditions under which the damage due to missingness
 can be reduced.  This theory has also resulted in
a number of performance guarantees when data obey certain
statistical conditions.  However, these conditions are rather strong, and
extremely hard to ascertain in real world problems. 
 \cite{Rubin2014}(page 22), summarize the state of
the art by observing: 
 ``\textit{essentially all the literature on multivariate incomplete data assumes that the data
are Missing At Random (MAR)}". 
Indeed, popular estimation  methods for missing data such as Maximum Likelihood based techniques \citep{dempster1977maximum} and Multiple Imputation \citep{rubin1978multiple} require MAR assumption to guarantee convergence to consistent estimates. 
%
Furthermore, it is almost impossible for a practicing statistician to decide whether the MAR condition holds in a given  problem. The literature on data that go beyond MAR is quite limited, and lacks systematic methodology for computing consistent estimates when such exist. 
 Some examples include \cite{fitzmaurice2008generalized}, \cite{carpenter2014missing}, \cite{robins2000robust} and \cite{scharfstein1999adjusting}.

Recent years have witnessed
a growing interest in analysing missing data  
using graphical models to encode
assumptions about the reasons for missingness.
This development is natural since graphical models
provide efficient representation of conditional independencies implied by modeling assumptions.  
Earlier papers in this development are \cite{daniel2012using} who provided sufficient criteria under which consistent estimates can be computed from complete cases (i.e. samples in which all variables are fully observed).
\cite{felix2013} (and later on \cite{felix2015}) developed techniques that guide the selection of auxiliary variables to improve estimability from incomplete data.
In machine learning, particularly while estimating
parameters of Bayesian Networks, graphical models have long been used as a tool when dealing with missing data (\cite{darwiche2009modeling}). 

\begin{table*}[t]
	\centering
		\caption{\textbf{Highlights of Major Results}}
		\newlength\q
\setlength\q{\dimexpr 1\textwidth -2\tabcolsep}
\noindent
	\begin{tabular}{p{\q}}
	\toprule
		\textbf{Criteria and procedures for recovering statistical and causal parameters from missing data}
	\\
	\midrule
		1. We provide methods for recovering conditional distributions in the presence of latent variables.\\
		2. We demonstrate the feasibility of recovering joint distribution in cases where variables cause their own missingness.\\
		3. We identify problems for which recoverability is infeasible.\\
					\hline
	\end{tabular}
	\label{tab:highlights1}
\end{table*}

\begin{table*}[t]
	\centering
		\newlength\qo
\setlength\qo{\dimexpr 1\textwidth -2\tabcolsep}
\noindent
	\begin{tabular}{p{\qo}}
	\toprule
		\textbf{Tests for challenging compatibility of model with observed data}
	\\
	\midrule
		1. We establish general criteria for testing conditional independence claims.\\
		2. We devise tests for MAR (Missing at Random) models.\\
		3. We identify dependence claims that defy testability.\\
					\hline
	\end{tabular}
	\label{tab:highlights2}
\end{table*}

In this paper we review the contributions of graphical models to
missing data research and emphasize three main aspects: (1) Transparency (2) Recoverability (consistent estimation) and (3) Testability. The main results of the paper are highlighted in table \ref{tab:highlights1}.

\paragraph{Transparency} 
Consider a practicing statistician  who has acquired a
statistical package that handles missing data and would like to
know whether the problem at hand meets
the requirements of the software.
As noted by \cite{Rubin2014} (see appendix \ref{app:rubin}) and
many others such as \cite{rhoads2012problems} and
\cite{balakrishnan2010methods}, almost all available
software packages implicitly assume that data
fall under two categories: MCAR (Missing Completely At Random) or MAR (formally defined in section
\ref{sec:mechanism}).
Failing this assumption, there is no guarantee that
estimates produced by current software will be
less biased than those produced by complete case analysis. 
Consequently, it
is essential for the user to decide if the type of
missingness present in the data is compatible with
the requirements of MCAR or MAR.

Prior to the advent of graphical models, no tool was
available to assist in this decision, since the
independence conditions that define MCAR or MAR
are neither visible in the data, nor in
a mathematical model that a researcher can consult to verify those conditions.
We will show how graphical models enable an efficient
and transparent classification of the missingness
mechanism.  In particular, the question of whether
the data fall into the  MCAR or MAR categories
can be answered by mere inspection of the graph structure.
 In addition, we will show how graphs facilitate 
a more refined, query-specific taxonomy of missingness
in MNAR (Missing Not At Random) problems.

The transparency associated with graphical models 
stems from three factors. First, graphs excel in
encoding and detecting conditional independence relations,
far exceeding the capacity of human intuition. Second,
all assumptions are encoded causally,
mirroring the way researchers store qualitative
scientific knowledge;
direct judgments of conditional independencies are not
required, since these  can be read off the structure of the graph.
Finally, the ultimate aim of all assumptions is to
encode ``the reasons for missingness" which is a causal,
not a statistical concept. Thus, even when our target
parameter is purely statistical, say a
regression coefficient, causal modeling is still
needed for encoding the ``process that causes missing data"
(\cite{Rubin_missing}).

\paragraph{Recoverability (Consistent Estimation)}
Recoverability (to be defined formally in Section \ref{sec:recoverability}) refers
to the task of determining, from an assumed model,
whether any method exists that produces a consistent
estimate of a desired parameter and, if so, how.
If the answer is negative, then an inconsistent estimate
should be expected even with large samples, and no algorithm,
however smart, can yield a consistent estimate.
On the other hand, if the answer  is affirmative
then there exists a procedure that can exploit the
features of the problem to produce consistent estimates.
If the problem is MAR or MCAR, standard missing data software can be used to obtain
consistent estimates. But if a recoverable problem is MNAR,
the user would do well to discard standard software and
resort to an estimator based on graphical analysis.
In Section \ref{sec:recoverability} of this paper we present
several methods of deriving consistent estimators for both statistical and causal parameters.

The general question of recoverability, to the best
of our knowledge, has not received due attention in the literature.
The notion that some parameters cannot be
estimated by any method whatsoever while others can,
still resides in an unchartered territory.
We will show in Section 3 that most MNAR problems
exhibit this dichotomy. That is, problems for which it is impossible
to properly impute all missing values in the data, would still
permit the consistent estimation of some parameters
of interest.  More importantly, the estimable parameters can often be identified directly
from the structure of the graph.

\paragraph{Testability}
Testability asks whether it is possible to tell
if any of the model's assumptions is incompatible with the
available data (corrupted by missingness).
Such compatibility tests under missingness are hard to come by and the
few tests reported in the literature are mostly 
limited to  MCAR \citep{little1988test}. As stated in \cite{allison2003missing}, 
``\textit{Worse still, there is no empirical way to discriminate one nonignorable model from another (or from the ignorable model).}''. In section
\ref{sec:testability} we will show that remarkably, discrimination is feasible; MAR problems do have a simple set of testable implications and MNAR problems can often be tested depending on their graph structures. 

In summary, although mainstream statistical analysis of missing data problems
has made impressive
progress in the past few decades, it left key problem areas
relatively unexplored, especially those touching on
transparency, estimability and testability. This paper
casts missing data problems in the language of causal graphs
and shows how this representation facilitates solutions
to pending problems. In particular, we show how the
MCAR, MAR, MNAR taxonomy becomes transparent in the
graphical language, how the estimability of a needed parameter
can be determined from the graph structure, what
estimators would guarantee consistent estimates, and
what modeling assumptions lend themselves to empirical
scrutiny.

\section{Graphical Models for Missing Data: Missingness Graphs (m-graphs)}
\begin{figure}[h]
	\centering
	\includegraphics[scale=0.8]{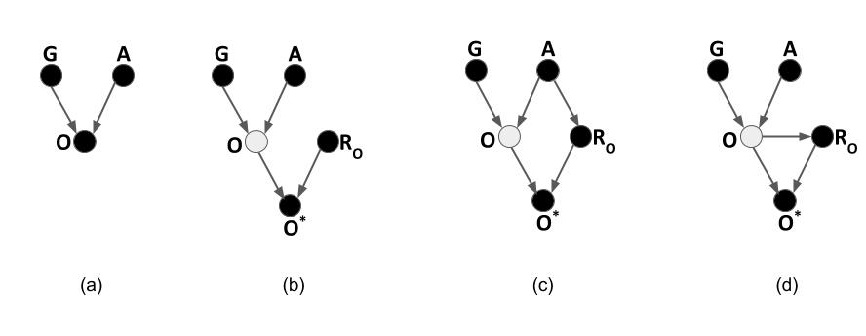}
	\caption{(a)causal graph under no missingness (b), (c) \& (d) m-graphs modeling distinct missingness processes.}
	\label{fig:mgraph}
\end{figure}
The following example, inspired by \cite{Rubin2} (example-1.6, page 8), describes how graphical models can be used to explicitly model the missingness process and encode the underlying causal and statistical assumptions. Consider a study conducted in a school that measured three (discrete) variables: Age (A), Gender (G) and Obesity (O).

\textbf{No Missingness }If all three variables are completely recorded, then there is no missingness. The causal graph\footnote{For a gentle introduction to causal graphical models
	see \cite{elwert2013graphical, lauritzen2001causal}, sections 1.2 and 11.1.2 in  \cite{pearl}.} depicting the interrelations between variables is shown in Figure \ref{fig:mgraph} (a). Nodes correspond to variables and edges indicate the existence of a causal relationship between pairs of nodes they connect. The value of a child node is a (stochastic) function of the values of its parent nodes. i.e. Obesity is a (stochastic) function of Age and Gender. The absence of an edge between Age and Gender indicates that $A$ and $G$ are independent, denoted by $A \compos G$. 

\begin{table}[h]
	\centering
	\caption{Missing dataset in which Age and Gender are fully observed and Obesity is partially observed.}
	\begin{tabular}{|c|c|c|c|c|}
		\hline
		$\#$ &Age & Gender & Obesity$^*$& $R_O$\\
		\hline
		1 & 16 & F & Obese&0\\
				2 & 15 & F & $m$ &1\\
				3 & 		15 & M & $m$&1\\
							4&	14 & F & Not Obese&0\\
								5&		13 & M & Not Obese&0\\
					6&	15 & M & Obese&0\\
						7&		14 & F & Obese&0\\
								\hline
	\end{tabular}
\label{tab:dataset}
\end{table}

\textbf{Representing Missingness} Assume that Age and Gender are are fully observed since they can be obtained from school records. Obesity however is corrupted by missing values due to some students not revealing their weight.  When the value of $O$ is missing we get an empty measurement which we designate by $m$. Table \ref{tab:dataset} exemplifies a missing dataset. The missingness process can be modelled using a proxy variable Obesity$^* (O^*)$ whose values are determined by Obesity and its missingness mechanism $R_O$.
\begin{equation}    
O^*=f (R_{O},O)  = \left\{
\begin{array}{l l} 
O & \quad \text{if }R_{O}=0\nonumber \\
m & \quad \text{if }R_{O}=1\\
\end{array} \right. 
\end{equation} 
$R_O$ governs the masking and unmasking of Obesity.   When $R_O=1$ the value of obesity is concealed i.e. $O^*$ assumes the values $m$ as shown in samples 2 and 3 in table \ref{tab:dataset}. When $R_O=0$, the true value of obesity is revealed i.e.  $O^*$ assumes the underlying value of Obesity as shown in samples 1, 4, 5, 6 and 7 in table \ref{tab:dataset}. 

Missingness can be caused by random processes or can depend on other variables in the dataset. An example of random missingness is students \textit{accidentally losing} their questionnaires. This is depicted in figure \ref{fig:mgraph} (b) by the absence of parent nodes for $R_O$.  Teenagers rebelling and not reporting their weight is an example of missingness  caused by a fully observed variable. This is depicted in figure \ref{fig:mgraph} (c) by an edge between $A$ and $R_O$. Partially observed variables can be causes of missingness as well. For instance consider obese students who are embarrassed of their obesity and hence reluctant to reveal their weight.  This is depicted in figure \ref{fig:mgraph} (d) by an edge between $O$ and $R_O$  indicating the $O$ is the cause of its own missingness. 

The following subsection formally introduces missingness graphs (m-graphs) as discussed in \cite{mohan2013}. 
\subsection{Missingness Graphs: Notations and Terminology} \label{sec:m-graph}
Let $G (\mathbf{V},E) $ be the causal DAG where $\mathbf{V}$ is the set of nodes and $E$ is the set of edges. Nodes in the graph  correspond to variables in the data set and are partitioned into five categories, i.e.  \begin{center}
	$\mathbf{V}=V_o \cup V_m\cup U \cup V^* \cup R$
\end{center}

$V_o $ is the set of variables that are observed in all records in the population and $V_m $ is the set of variables that are missing in at least one record.  Variable $X$ is termed as \textit{fully observed} if $X \in V_o$ and \textit{partially observed} if $X \in V_m$.  
$R_{v_i}$ and $V_i^*$ are two variables associated with every partially observed variable, where $V_i^*$ is a proxy variable that is actually observed, and $R_{v_i}$ represents the status of the causal mechanism responsible for the missingness of $V_i^*$; formally,
\begin{equation}    
v_i^*=f (r_{v_i},v_i)  = \left\{
\begin{array}{l l} \label{eq:miss1}
v_i & \quad \text{if }r_{v_i}=0\\
m & \quad \text{if }r_{v_i}=1\\
\end{array} \right. 
\end{equation}
$V^*$ is the set of all proxy variables and $\mathbf{R}$ is the set of all causal mechanisms that are responsible for missingness. $U$ is the set of unobserved nodes, also called latent variables. 
Unless stated otherwise it is assumed that no variable in $V_o \cup V_m \cup U$ is a child of an $R$ variable.  
Two nodes $X$ and $Y$ can be connected by a directed edge i.e. $X \to Y$, indicating that $X$ is a cause of $Y$, or by a bi-directed edge $X<\hspace*{-2mm}\textendash \textendash \hspace*{-2mm}>Y$ denoting the existence of a $U$ variable that is a parent of both $X$ and $Y$. 

We call this graphical representation a \textbf{Missingness Graph} (or $m$-graph). Figure \ref{fig:mgraph} exemplifies three m-graphs 
in which $V_o=\{A,G\}$, $V_m=\{O\}$, $V^*=\{O^*\}$, $U=\emptyset$ and $R=\{R_O\}$. Proxy variables may not always be explicitly shown in m-graphs in order to keep the figures simple and clear. 
The missing data distribution, $P(V^*,V_o,R)$ is referred to as the \textit{observed-data distribution}  and the distribution that we would have obtained had there been no missingness, $P(V_o,V_m,R)$ is called as the \textit{underlying distribution}. Conditional Independencies are read off the graph using the d-separation\footnote{For an introduction to d-separation see, http://bayes.cs.ucla.edu/BOOK-2K/d-sep.html and  http://www.dagitty.net/learn/dsep/index.html} criterion \citep{pearl}. For example, Figure \ref{fig:mgraph} (c)
depicts the independence $R_O \compos O|A$ but not $R_O \compos G|O$. %

\subsection{Classification of Missing Data Problems based on Missingness Mechanism}
\label{sec:mechanism}

\cite{Rubin_missing} 
 classified missing data into three categories: Missing Completely At Random (MCAR), Missing At Random (MAR) and Missing Not At Random (MNAR)  based on the statistical dependencies between the missingness mechanisms ($R$ variables) and the variables in the dataset $(V_m,V_o)$. We capture the essence of this categorization in graphical terms below.
\begin{enumerate}
	\item Data are MCAR if $V_m \cup V_o \cup U \compos R$ holds in the m-graph. In  words,  missingness occurs completely at random and is entirely independent of both the observed and the partially observed variables. This condition can be easily identified in an m-graph by the absence of edges between the $R$ variables and variables in $V_o \cup V_m$. 
	\item Data are MAR if  $V_m \cup U \compos R|V_o$ holds in the m-graph. In words, conditional on the fully observed variables $V_o$, missingness occurs
	at random. In graphical terms, MAR holds if (i) no edges exist between an $R$ variable and any partially observed variable and (ii) no bidirected edge exists between an $R$ variable and a fully observed variable.  MCAR implies MAR, ergo all estimation techniques applicable to MAR can be safely  applied to MCAR. 
	\item Data that are not MAR or MCAR fall under the MNAR category. 
\end{enumerate}
m-graphs in figure \ref{fig:mgraph} (b), (c) and (d) are typical examples of MCAR, MAR and MNAR categories, respectively. Notice the ease with which the three categories can be identified. Once the user lays out the interrelationships between the variables in the problem, the classification is purely mechanical. 

\subsubsection{Missing At Random: A Brief Discussion} The original classification used in \cite{Rubin_missing} is very similar to the one defined in the preceding paragraphs. The main distinction rests on the fact that MAR defined in \cite{Rubin_missing} (which we call \textit{Rubin-MAR}) is defined in terms of conditional independencies between events where as that in this paper (referred to as MAR) is defined in terms of conditional independencies between variables. Clearly, we can have the former without the latter, in practice though it is rare that scientific knowledge can be articulated in terms of event based independencies that are not implied by variable based independencies.  

Over the years the classification proposed in \cite{Rubin_missing}
has been criticized both for its nomenclature and its opacity.
Several authors noted that \textbf{MAR is
a misnomer} (\cite{scheffer2002dealing,  peters2002primer,
	meyers2006applied, graham2009missing}) noting that randomness in this class is critically conditioned on observed data.  

However, the \textbf{opacity of the assumptions} underlying Rubin's
MAR presents a more serious problem.
Clearly, a researcher would find it cognitively taxing, if
not impossible to even decide if any of these independence assumptions is  
reasonable. This, together
with the fact that
Rubin-MAR is untestable (\cite{impute})  motivates
the variable-based taxonomy presented above. \cite{seaman2013meant} and \cite{doretti2018missing} provide another taxonomy and a different perspective on Rubin-MAR.

Nonetheless, Rubin-MAR has an interesting theoretical property:  
It is the  weakest simple condition under
which the process that causes
missingness can be ignored while still making correct inferences about the data \citep{Rubin_missing}.
It was probably this theoretical result that
changed missing data practices in the 1970's.
The popular practice prior to 1976 
was to assume that missingness was caused
totally at random (\cite{gleason1975proposal,
haitovsky1968missing}). With Rubin's identification of the
MAR condition as sufficient for drawing correct inferences,
MAR became the main focus of attention in the statistical literature.

Estimation procedures such as Multiple Imputation were developed and implemented with
MAR assumptions in mind, and popular textbooks were authored
exclusively on MAR and its simplified versions \citep{grahambook}.
In the absence of recognizable criterion for MAR, some
authors have devised heuristics invoking auxiliary
variables, to increase the chance of achieveing MAR
\citep{collins2001comparison}. Others have warned against indiscriminate
inclusion of such variables \citep{felix2013,felix2015}.
These difficulties have engendered a culture with a tendency to
blindly assume MAR, with the consequence that
the more commonly occurring MNAR class of problems
remains relatively unexplored
\citep{resseguier2011sensitivity, adams2007researching, osborne2012best, osborne2014best,
	sverdlov2015modern,  van2016incremental}. 

In his seminal paper \citep{Rubin_missing} Rubin recommended that researchers explicitly model the missingness process: 
\begin{figure}[H]
	\centering
	\includegraphics[scale=1.2]{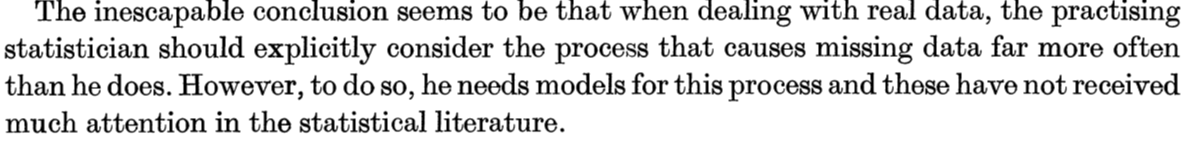}
	\caption{Quote from \cite{Rubin_missing}}
	\label{fig:rubin}
\end{figure}

This recommendation invites in fact the graphical tools described in this paper, for they encourage investigators to 
model the details of the missingness process rather than blindly assume MAR. These tools have further enabled researchers
  to extend the
analysis of estimation  to the vast class
of MNAR problems.

In the next section we discuss how graphical models
accomplish these tasks.

\section{Recoverability}
\label{sec:recoverability}
Recoverability\footnote{The term identifiability is sometimes used in lieu of recoverability. We prefer using recoverability over \textit{identifiability} since the latter is strongly associated with
causal effects, while the former is a broader concept, applicable
to statistical relationships as well. See section \ref{sec:causal}. } addresses the basic question of whether a quantity/parameter of interest can be estimated from incomplete data \textit{as if} no missingness took place, that is, the desired quantity can be estimated consistently from the available (incomplete) data. This amounts to expressing the target quantity $Q$ in terms of the observed-data distribution $P(V^*,V_O,R)$.  Typical target quantities that shall be considered are
conditional/joint distributions and conditional causal effects.
\begin{definition}[Recoverability of target quantity $Q$]  Let $A$ denote the set of assumptions about the data generation process and let $Q$ be any functional of the underlying distribution $P(V_m,V_O,R)$. $Q$ is recoverable if there exists a procedure that computes a  consistent estimate of $Q$ for all strictly positive  observed-data distributions $P(V^*,V_o,R)$ that may be generated under $A$.\footnote{This definition is more operational than the standard definition of identifiability for it states explicitly what is achievable under recoverability and more importantly, what problems may occur under non-recoverability. } 
\end{definition}
Since we encode all assumptions in the structure of the m-graph
$G$, recoverability becomes a property of the pair \{$Q, G$\}, and not of the data.
We restrict the definition above to strictly positive observed-data distributions, $P(V^*,V_o,R)$ except for  instances of zero probabilities as specified in equation \ref{eq:miss1}.  The reason for this restriction  can be understood as the need for observing some unmasked cases for all combinations of variables, otherwise, masked cases can be arbitrary.
We note however that recoverability is sometimes feasible even when strict positivity does not hold (\cite{mohan2013}, definition 5 in appendix).

We now demonstrate how a joint distribution is recovered given MAR data.
\begin{example}
	Consider the problem of recovering the joint distribution given the $m$-graph in Fig.\ \ref{fig:mgraph} (c) and dataset in table \ref{tab:mardata}.  Let it be the case that 15-18 year olds were reluctant to reveal their weight, thereby making $O$ a partially observed variable i.e. $ V_m=\{O\}$ and $V_o=\{G,A\}$.  This is a typical case of MAR missingness, since the cause of missingness is the fully observed variable: Age. The following three steps detail the recovery procedure. 
	\begin{align*}
	\intertext{1. Factorization: The joint distribution may be factorized as:}
	P(G,O,A)&=P(G,O|A)P(A)\\
	\intertext{2. Transformation into observables:  $G$ implies the conditional independence $(G,O) \compos R_O |A$ since $A$ d-separates $(G,O)$ from $R_O$. Thus,}
	P(G,O,A)&=P(G,O|A,R_O=0)P(A) \\
	\intertext{3. Conversion of partially observed variables into proxy variables: 	$R_O=0$ implies $O^*=O$ (by eq \ref{eq:miss1}).  Therefore,}
	P(G,O,A)&=P(G,O^*|A,R_O=0)P(A) \numberthis \label{eq:mar}
	\end{align*}
	The RHS of Eq. (\ref{eq:mar}) is expressed in terms of variables  in the observed-data distribution. Therefore, $P (G,A,O) $ can be consistently estimated (i.e. recovered) from the available data. The recovered joint distribution is shown in table \ref{tab:recovered}. 
	\label{ex:mar}
\end{example}
\begin{table}[h]
	\caption{observed-data Distribution $P(G,A,O^*,R_O)$ where Gender $(G)$ and Age $(A)$ are fully observed, Obesity $O$ is corrupted by missing values and Obesity's  proxy $(O^*)$ is observed in its place. Age is partitioned into three groups:  $[10-13), [13-15), [15-18)$. Gender and Obesity are binary variables and can take values Male (M) and Female (F), and Yes (Y) and No (N), respectively. The probabilities $p_1, p_2,..p_{18}$ stand for the (asymptotic) frequencies of the samples falling in the 18 cells ($G, A, O^*, R_O $). } 	
\begin{minipage}[t]{.45\textwidth}
	\begin{tabular}{|c|c|c|c|c|}
		\hline
		$G$  & $A$  & $O^*$ & $R_O$ & $P(G,A,O^*,R_O)$ \\
		\hline
		M & $10-13$ & Y & $0$ & $p_1$\\
		M & $13-15$ & Y & $0$ & $p_2$\\
		M & $15-18$ & Y & $0$ & $p_3$\\
		M & $10-13$ & N & $0$ & $p_4$\\
		M & $13-15$ & N & $0$ & $p_5$\\
		M & $15-18$ & N & $0$ & $p_6$\\
		F & $10-13$ & Y & $0$ & $p_7$\\
		F & $13-15$ & Y & $0$ & $p_8$\\
		F & $15-18$ & Y & $0$ & $p_9$\\
		\hline
	\end{tabular}
\end{minipage}
\hfill
\noindent
\begin{minipage}[t]{.45\textwidth}
\begin{tabular}{|c|c|c|c|c|}
	\hline
	$G$  & $A$  & $O^*$ & $R_O$ & $P(G,A,O^*,R_O)$ \\
	\hline
	F & $10-13$ & N & $0$ & $p_{10}$\\
	F & $13-15$ & N & $0$ & $p_{11}$\\
	F & $15-18$ & N & $0$ & $p_{12}$\\
	M & $10-13$ & $m$ & $1$ & $p_{13}$\\
	M & $13-15$ & $m$ & $1$ & $p_{14}$\\
	M & $15-18$ & $m$ & $1$ & $p_{15}$\\
	F & $10-13$ & $m$ & $1$ & $p_{16}$\\
	F & $13-15$ & $m$ & $1$ & $p_{17}$\\
	F & $15-18$ & $m$ & $1$ & $p_{18}$\\
			\hline
	\end{tabular}	
\end{minipage}
\label{tab:mardata}
\end{table}
\begin{table}[h]
	\caption{Recovered joint distribution corresponding to dataset in table \ref{tab:mardata} and m-graph in figure \ref{fig:mgraph}(c)}
	\centering
	\begin{minipage}[t]{.5\textwidth}
	\begin{tabular}{|c|c|c|c|}
	\hline
	$G$  & $A$  & $O$ &  $P(G,O,A)$ \\
	\hline	
	M & $10-13$ & Y  & $\frac{p_1*(p_1+p_4+p_7+p_{10}+p_{13}+p_{16})}{p_1+p_4+p_7+p_{10}}$\\
	M & $13-15$ & Y  & $\frac{p_2*(p_2+p_5+p_8+p_{11}+p_{14}+p_{17})}{p_2+p_5+p_8+p_{11}}$\\
	M & $15-18$ & Y  & $\frac{p_3*(p_3+p_6+p_9+p_{12}+p_{15}+p_{18})}{p_3+p_6+p_9+p_{12}}$\\
	M & $10-13$ & N  & $\frac{p_4*(p_1+p_4+p_7+p_{10}+p_{13}+p_{16})}{p_1+p_4+p_7+p_{10}}$\\
	M & $13-15$ & N  & $\frac{p_5*(p_2+p_5+p_8+p_{11}+p_{14}+p_{17})}{p_2+p_5+p_8+p_{11}}$\\
	M & $15-18$ & N  & $\frac{p_6*(p_3+p_6+p_9+p_{12}+p_{15}+p_{18})}{p_3+p_6+p_9+p_{12}}$\\
	\hline
	\end{tabular}
	\end{minipage}
\begin{minipage}[t]{.48    \textwidth}
	\begin{tabular}{|c|c|c|c|}
		\hline
		$G$  & $A$  & $O$ &  $P(G,O,A)$ \\
		\hline	
		F & $10-13$ & Y  & $\frac{p_7*(p_1+p_4+p_7+p_{10}+p_{13}+p_{16})}{p_1+p_4+p_7+p_{10}}$\\
		F & $13-15$ & Y  & $\frac{p_8*(p_2+p_5+p_8+p_{11}+p_{14}+p_{17})}{p_2+p_5+p_8+p_{11}}$\\
		F & $15-18$ & Y  & $\frac{p_9*(p_3+p_6+p_9+p_{12}+p_{15}+p_{18})}{p_3+p_6+p_9+p_{12}}$\\
		F & $10-13$ & N  & $\frac{p_{10}*(p_1+p_4+p_7+p_{10}+p_{13}+p_{16})}{p_1+p_4+p_7+p_{10}}$\\
		F & $13-15$ & N  & $\frac{p_{11}*(p_2+p_5+p_8+p_{11}+p_{14}+p_{17})}{p_2+p_5+p_8+p_{11}}$\\
		F & $15-18$ & N  & $\frac{p_{12}*(p_3+p_6+p_9+p_{12}+p_{15}+p_{18})}{p_3+p_6+p_9+p_{12}}$\\
		\hline
	\end{tabular}
\end{minipage}
\label{tab:recovered}
\end{table}
	Note that samples in which obesity is missing are not discarded but are used instead to update the weights $p_1,..p_{12}$ of the cells in which obesity is has a definite value. This can be seen by the presence of probabilities $p_{13},...p_{18}$ in table \ref{tab:recovered} and the fact that samples with missing values have been utilized to estimate prior probability $P(A)$ in equation \ref{eq:mar}. Note also that the joint distribution permits an alternative decomposition: 
	\begin{align*}
	P(G,O,A) &= P(O|A,G) P(A,G)\\
	&= P(O^*|A,G,R_O=0) P(A,G)	
	\end{align*}
	The equation above licenses a different estimation procedure whereby $P(A,G)$ is estimated from all samples, including those in which obesity is missing, and only the estimation of $P(O^*|A,G,R_O=0)$ is restricted to the complete samples. 
	The efficiency of various decompositions are analysed in \cite{mohan_learning, closedform}. 
	
	Finally we observe that for the MCAR m-graph in figure \ref{fig:mgraph} (b), a wider spectrum of decompositions is applicable, including:
	\begin{align*}
	P(G,O,A) &= P(O,A,G|R_O=0) \\
	&=	P(O^*,A,G|R_O=0)
	\end{align*}
	The equation above licenses the estimation of the joint distribution using only those samples in which obesity is observed. This  estimation procedure, called listwise deletion or complete-case analysis \citep{Rubin2},  would usually result in wastage of data and lower quality of estimate, especially when the number of samples corrupted by missingness is high. Considerations of estimation efficiency should therefore be applied once we explicate the spectrum of options licensed by the m-graph. 

		A completely different behavior will be encountered
		in the model of \ref{fig:mgraph} (d) which, as we have noted,
		belong to the MNAR category. Here, the arrow $O \to R_O$
		would prevent us from executing step 2  of the estimation
		procedure, that is, transforming $P(G,O,A)$ into an expression
		involving solely observed variables. We can in fact show
		that in this example the joint distribution is
		nonrecoverable. That is, regardless of
		how large the sample or how clever the imputation, 
		no algorithm exists that produces consistent estimate of
		P(G,O,A).

 The possibility of encountering non-recoverability is not discussed as often as it ought to be in mainstream missing data literature mostly because the MAR assumption is either taken for granted \citep{Pfeffermann} or thought of as a good approximation for MNAR \citep{chang2011modern}. Consequently it is often presumed that the maximum likelihood method can deliver a consistent estimate of any desired parameter.  While it is true for MAR, it is certainly not true in cases for which we can prove non-recoverability, and requires model-based analysis for MNAR. 		

\begin{remark}
 Observe that equation \ref{eq:mar} yields an \textbf{estimand} for the query, $P(G,O,A)$, as opposed to an \textit{estimator}. An estimand is a functional of the observed-data distribution, $P(V^*,R,V_o)$, whereas an estimator is a rule detailing how to calculate the estimate from measurements in the sample. Our estimands naturally give rise to a closed form estimator, for instance, the estimator corresponding to the estimand in equation \ref{eq:mar} is: \\ $\frac{\#(G=g,O^*=o,A=a,R_O=0)}{\#(A=a,R_O=0)}\frac{\#(A=a)}{N}$, where $N$ is the total number of samples collected and $\#(X_1=x_1,X_2=x_2,...X_j=x_j)$ is the frequency of the event $x_1,x_2,...x_j$. 
Algorithms  inspired by such closed form estimation techniques 
were shown in \cite{mohan_learning}, to outperform conventional methods such as EM computationally, for instance by scaling to networks where it is intractable to run even one iteration of EM. Such algorithms are indispensable for large scale and big data learning tasks in machine learning and artificial intelligence for which EM is not a viable option.
\end{remark}

\begin{figure}[h]
	\centering
	\includegraphics[scale=0.8]{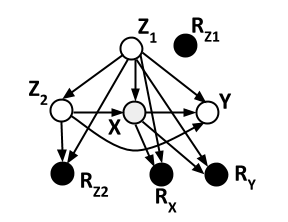}
	\caption{m-graphs depicting MNAR in which $P(Y|X,Z_1,Z_2)$ is recoverable. Proxy variables have not been explicitly portrayed as stated in section \ref{sec:m-graph}. }
	\label{fig:reviewersExample}
\end{figure}

\paragraph{Recovering from Complete \& Available cases} Traditionally there has been great interest in \textit{complete case analysis } primarily due to its simplicity and ease of applicability. 
However, it results in a large wastage of data and
a more economical version of it, called \textit{available case analysis}
would generally be more desirable.
The former retains only samples in which
variables in the entire dataset are observed, whereas the latter retains all samples in which the variables in the query
are observed.  
Sufficient criterion for recovering
conditional distributions from complete cases as well as available cases is widely discussed in literature( \cite{bartlett2014improving, Rubin2, white2010bias}) 
and we state them in the form of a corollary below:
\begin{cor}
(a) Given m-graph $G$, $P(X|Y)$ is recoverable from
complete cases if $X \compos R|Y$ holds in $G$ where $R$ is the set of
all missingness mechanisms. \\
(b) Given m-graph $G$, $P(X|Y)$ is
recoverable from available cases if $X \compos (Rx,Ry)|Y$ holds in $G$.
\end{cor}
In figure \ref{fig:reviewersExample} for example, we see that $Z_1 \compos R_{Z_1}$ holds but $Z \compos R_x$ does not. Therefore $P(Z_1)$
is recoverable from available cases but not complete cases.

A generic example for recoverability under MNAR is presented below. 

\begin{example}[Recoverability in MNAR m-graphs] 
    Consider the m-graph $G$ in figure \ref{fig:reviewersExample} where all variables are subject to missingness. $Y$ is the outcome of interest, $X $ the exposure of interest and $Z_1$ and $Z_2$ are baseline covariates.  The target parameter is $P(Y|X,Z_1,Z_2)$,  the regression of $Y$ on $X$ given both  baseline covariates.\\ Since $Y \compos (R_X,R_Y,R_{Z_1},R_{Z_2})|(X,Z_1,Z_2)$ in $G$, $ P(Y|X,Z_1,Z_2)$ can be recovered as:
    \begin{align}
        P(Y|X,Z_1,Z_2) &= P(Y|(X,Z_1,Z_2, R_X=0,R_Y=0,R_{Z_1}=0,R_{Z_2}=0)) \nonumber \\
        &= P(Y^*|(X^*,Z_1^*,Z_2^*, R_X=0,R_Y=0,R_{Z_1}=0,R_{Z_2}=0))  \mbox{( Using eq \ref{eq:miss1})} \nonumber
    \end{align}
    Note that despite the fact that all variables are subject to missingness and missingness is highly dependent on partially observed variables the graph nevertheless licenses the estimation of the target parameter from samples in which all variables are observed.
\end{example}


In the following subsection we define the notion of Ordered
factorization which leads to a criterion for sequentially
recovering conditional probability distributions (\cite{mohan2013,mohan2014_missing}).

\subsection{Recovery by Sequential Factorization}

\begin{figure}[h]
	\centering
	\includegraphics[scale=1.5]{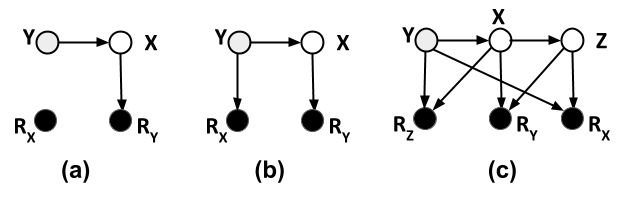}
	\caption{m-graphs from which joint and/or conditional distributions can be recovered using various factorizations. }
	\label{fig:entangled}
\end{figure}

\begin{definition}[Ordered factorization of $P(Y|Z)$] 
	Let $Y_1<Y_2 < \ldots < Y_k$ be an ordered set of all variables in $Y$, $1 \le i\le |Y|=k$ and $X_i \subseteq \{Y_{i+1}, \ldots, Y_n\}\cup Z$. 	
	 Ordered factorization of $P(Y|Z)$  is the product of conditional probabilities i.e. $P(Y|Z) =\prod_i P(Y_i|X_i)$, 
	 such that $X_i$ is a minimal set for which  $Y_i\compos (\{Y_{i+1}, \ldots, Y_n\}\setminus X_i)|X_i$ holds.
\end{definition}
The following theorem presents a sufficient condition for recovering conditional distributions of the form $P(Y|X)$ where $\{Y,X\}\subseteq V_m\cup V_o$.
\begin{thm} 
	Given an m-graph $G$ and a observed-data distribution $P(V^*,V_o,R)$, a target quantity $Q$ is recoverable if $Q$ can be decomposed into
	an ordered factorization, or a sum of such factorizations,
	such that every factor $Q_i=P(Y_i | X_i)$ satisfies
	$Y_i \compos (R_{y_i}, R_{x_i}) | X_i$. 	Then, each $Q_i$ may be recovered as $P(Y^*_i|X^*_i,R_{Y_i}=0,R_{X_i}=0)$.
	\label{thm:seq}
\end{thm}
An ordered factorization that satisfies theorem \ref{thm:seq} is called as an \textit{admissible factorization}.
\begin{example}
Consider the problem of recovering $P(X,Y)$ given $G$, the m-graph in figure \ref{fig:entangled} (a). $G$ depicts an MNAR problem since missingness in $Y$ is caused by the partially observed variable $X$. The factorization $P(Y|X)P(X)$ is admissible  since both $Y \compos R_x,R_y|X$ and  $X \compos R_x$ hold in $G$. 
$P(X,Y)$ can thus be recovered using theorem \ref{thm:seq} as $P(Y^*|X^*,R_x=0,R_y=0)P(X^*|R_x=0)$.  Here, complete cases are used to estimate $P(Y|X)$
and all samples including those in which Y is missing are used to estimate $P(X)$.
Note that the decomposition $P(X|Y)P(Y)$ is not
admissible.
\end{example}

\begin{cor} Given an m-graph $G$ depicting MAR  joint distribution is recoverable in $G$ as $P(V_o,V_m)=P(V^*|V_o,R=0)P(V_o)$. 
	\label{cor:mar}
\end{cor}

\subsection{R Factorization}
\begin{example} 
	Consider the problem of recovering $Q=P(X,Y)$
	from the $m$-graph of Figure \ref{fig:entangled}(b). Interestingly, no ordered factorization over variables $X$ and $Y$ would satisfy the conditions
	of Theorem \ref{thm:seq}. To witness we write  
	$P(X,Y)=P(Y|X)P(X)$ and note that the graph does not
	permit us to augment any of the two terms with the
	necessary $R_x$ or $R_y$ terms; $X$ is independent of $R_x$ only
	if we condition on $Y$, which is partially observed, and $Y$
	is independent of $R_y$ only if we condition on $X$ which
	is also partially observed. This deadlock can be disentangled
	however using a non-conventional decomposition:
	\begin{align}
	Q & = P(X,Y) 
	 = P(X,Y)  \frac{P(R_x=0,R_y=0|X,Y)}{P(R_x=0,R_y=0|X,Y)}     \nonumber \\              
	&= \frac{P(R_x=0,R_y=0) P(X,Y|R_x=0,R_y=0)}{P(R_x=0|Y,R_y=0)P(R_y=0|X,R_x=0)}\nonumber
	\end{align}             
	where the denominator was obtained using the
	independencies $R_x \compos (X,R_y)|Y$ and \\ $R_y \compos (Y,R_x)|X$
	shown in the graph. The final expression below,
	\begin{align}
	P(X,Y)&= \frac{P(R_x=0,R_y=0) P(X^*,Y^*|R_x=0,R_y=0)}{P(R_x=0|Y^*,R_y=0)P(R_y=0|X^*,R_x=0)} \mbox{ (Using equation \ref{eq:miss1})}
	\label{ex5-eq}
	\end{align}
	which is in terms of variables in the observed-data distribution, renders
	$P(X,Y)$ recoverable.
	This example again shows that recovery is  feasible
	even when data are MNAR. 
	\label{ex5}
\end{example}
The following theorem (\cite{mohan2013,mohan2014_missing}) formalizes the recoverability scheme exemplified above.
\begin{thm}[Recoverability of the Joint $P(V)$] 
	Given a $m$-graph $G$ with no edges between $R$ variables 
	the necessary and sufficient condition for recovering
	 the joint distribution $P(V)$ is the absence of any variable
	$X \in V_m$ such that:\\
	1.  $X$ and $R_x$ are neighbors\\
	2. $X$ and $R_x$ are connected by a path in which all intermediate nodes are colliders\footnote{A variable is a collider on the path if the path enters and leaves
		the variable via arrowheads (a term suggested by the collision of causal forces
		at the variable) \citep{greenland2011causal}.} and elements of $V_m \cup V_o$.
	When recoverable, $P(V)$ is given by \begin{align}\label{eq-2013524946}
	P(v)=\frac{P(R=0,v)}{\prod_i P(R_i=0|Mb^o_{r_i},Mb^m_{r_i}, R_{Mb^m_{r_i}}=0)},
	\end{align} 
	where $Mb^o_{r_i}\subseteq V_o$ and $Mb^m_{r_i}\subseteq V_m$ are the markov blanket\footnote{ Markov blanket $Mb_X$ of variable $X$ is any set of variables such that $X$ is conditionally independent of all the other variables in the graph given $Mb_X$ \citep{pearl1988probabilistic}. } of $R_i$.
	\label{thm:r}
\end{thm}
The preceding theorem can be applied to immediately yield an  estimand for joint distribution. For instance, given the m-graphs in figure \ref{fig:entangled} (c), joint distribution can be recovered in one step yielding: 
 \begin{center}
 	$P(X,Y,Z)=\frac{P(X,Y,Z,R_x=0,R_y=0,R_z=0)}{P(R_x=0|Y,R_y=0,Z,R_z=0)P(R_y=0|X,R_x=0,Z,R_z=0)P(R_z=0|Y,R_y=0,X,R_x=0)}$
 \end{center} 
\begin{figure}[h]
	\centering
		\includegraphics[scale=1.5]{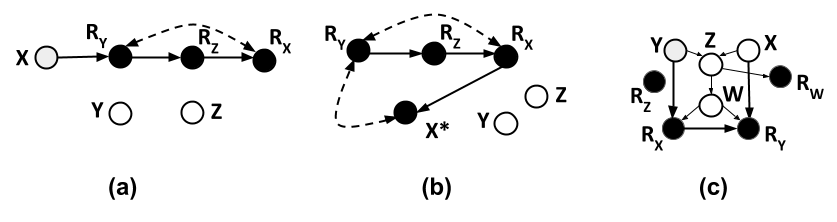}
	\caption{(a) \& (c) m-graphs from which conditional distributions can be recovered aided by intervention, (b) latent structure (\cite{pearl}, chapter 2) corresponding to m-graph in (a) when $X$ is treated as a latent variable. }
	\label{fig:inducing}
	\end{figure}
\subsection{Constraint Based Recoverability}
The recoverability procedures presented thus far relied entirely on conditional independencies that are read off the m-graph using d-separation criterion. Interestingly, recoverability can sometimes be accomplished by graphical patterns other than conditional independencies. These patterns represent distributional constraints which can be detected using mutilated versions of the m-graph. We describe below an example of constraint based recovery. 
\begin{example} Let $G$ be the m-graph in figure \ref{fig:inducing} (a) and let the query of interest be $P(X)$. The absence of a set that d-separates $X$ from $R_x$, makes it impossible to apply any of the techniques discussed previously. While it may be tempting to conclude that $P(X)$ is not recoverable, we prove otherwise by using the fact that 
	$X \compos R_x$ holds in the ratio distribution
	$\frac{P(X,R_y,R_z,R_x)}{P(R_z|R_y)}$.
	Such ratios are called interventional distributions and the resulting constraints are called Verma Constraints (\cite{inducing,tian2002testable}). The proof presented below employs the rules of do-calculus\footnote{For an introduction to do-calculus see, \cite{pearl2014external}, section 2.5 and \cite{koller2009probabilistic}}, to extract these constraints.
	\begin{align*}
	P(X) &= P(X|do(R_z=0)) \mbox{ (Rule-3 of do-calculus)}\\
	&= P(X|do(R_z=0),R_x=0)\mbox{ (Rule-1 of do-calculus)}\\
	&= P(X^*|do(R_z=0),R_x=0) \mbox{ (using equation \ref{eq:miss1})}\\
	&=\sum_{R_Y}P(X^*,R_Y|do(R_z=0),R_x=0) \numberthis \label{eq:condInterventional}
	\intertext{Note that the query of interest is now a function of $X^*$ and not $X$. Therefore the problem now amounts to identifying a conditional interventional distribution using the m-graph in figure \ref{fig:inducing}(b). A complete analysis of such problems is available in \cite{ilyabestpaper} which identifies the causal effect in eq \ref{eq:condInterventional} as: }
	P(X) &= \sum_{R_Y} P(X^*|R_Y,R_x=0,R_z=0)\frac{P(R_x=0|R_y,R_z=0)P(R_y)}{\sum_{R_Y}P(R_x=0|R_y,R_z=0)P(R_y)}\numberthis \label{eq:p(xy)}
	\end{align*}
In addition to $P(X)$, this graph also allows recovery of joint distribution  as shown below. 
$
P(X,Y,Z) = P(X)P(Y)P(Z)
$\\
{\normalsize
$
P(X,Y,Z) = \big(\sum_{R_Y} P(X^*|R_Y,R_x=0,R_z=0)\frac{P(R_x=0|R_y,R_z=0)P(R_y)}{\sum_{R_Y}P(R_x=0|R_y,R_z=0)P(R_y)}\big)\\ \hspace*{10cm} P(Y^*=Y|R_y=0) P(Z^*|R_z=0)
$
}

The decomposition in the first line uses $(X,Y)\compos Z$  and $X\compos Y$. Recoverability of $P(X)$ in the second line follows from equation \ref{eq:p(xy)}. Theorem \ref{thm:seq} can be applied to recover, $P(Y)$ and $P(Z)$, since $Y \compos R_Y$ and $Z \compos R_Z$. 
\end{example}
\begin{remark}
	In the preceding example  we were able to recover a joint distribution despite the fact that the distribution  $P(X,R_Y,R_x)$ is void of independencies. The ability to exploit such cases further underscores the need for graph based analysis. 
\end{remark}

The field of epidemiology has several impressive works dealing with coarsened data (\cite{gill1997coarsening, gill1997sequential}) and missing data (\cite{robins2000robust, robins1997non, robins2000sensitivity,li2013weighting}). Many among these  are along the lines of estimation (mainly of causal queries);  \cite{robins1994estimation} and \cite{rotnitzky1998semiparametric} deal with Inverse Probability Weighting based estimators, and  \cite{bang2005doubly} demonstrates the efficacy of Doubly Robust estimators using simulation studies. The recovery strategy of these existing works are different from that discussed in this paper with the main difference being that  these works proceed by intervening on the $R$ variable and thus converting the missing data problem into that of identification of causal effect. For example the problem of  recovering $P(X)$ is transformed into that of identifying the counterfactual query $P(X^*_{R_x=0})$ (which in our framework translates to identifying  $P(X^*|do(R_x=0))$) in the graph in which $X$ is treated as a latent variable. This technique while applicable in several cases is not general and may not always be relied upon to establish recoverability.  An example is the problem of recovering joint distribution $P(W,X,Y,Z)$ in figure \ref{fig:inducing} (c). In this case the equivalent causal query $P(W^*,X^*,Y^*,Z^*|do(R_x=0,R_y=0,R_w=0,R_z=0))$ is not identifiable in the graph in which $W,X,Y$ and $Z$ are treated as latent variables. The procedure for recovering  joint distribution from the m-graph in figure \ref{fig:inducing} (c) is presented in Appendix \ref{sec:complex}. 

\subsection{Overcoming Impediments to Recoverability}
This section focuses on MNAR problems that are not recoverable\footnote{Unless otherwise specified non-recoverability will assume joint distribution as a target and does not exclude recoverability of targets such as odds ratio (discussed in \cite{bartlett2015asymptotically}). }. One such problem is elucidated in the following example. 
\begin{example}
Consider a missing dataset comprising of a single variable, Income ($I$), obtained from a population in which the very rich and the very poor were reluctant to reveal their income. The underlying  process can be described as a variable causing its own missingness. The m-graph depicting this process is $I \to R_I$.  Obviously,  under these circumstances the true distribution over income, $P(I)$, cannot be computed error-free even if we were given infinitely many samples. 
\end{example}
The following theorem identifies graphical conditions that forbid recoverability of conditional probability distributions (\cite{mohan2014_missing}).
\begin{thm} Let $X \cup Y \subseteq V_m \cup V_o$ and $|X|=1$.
	$P(X|Y)$ is not recoverable if  either, $X$ and $R_{X}$ are neighbors or there exists a path from $X$ to $R_x$ such that all intermediate nodes are colliders and elements of $Y$. \label{thm:nonrecov}
\end{thm}

Quite surprisingly, it is sometimes possible to recover joint distributions given m-graphs with graphical structures stated in theorem \ref{thm:nonrecov} by jointly harnessing features of the data and m-graph. We exemplify such recovery with an example.
\begin{example}
	Consider the problem of recovering $P(Y,I)$ given the m-graph $G:$ $Y \to I \to R_I$, where $Y$ is a binary variable that denotes whether candidate has sufficient years of relevant work experience and $I$ indicates income. $I$ is also a binary variable and takes values high and low. $P(Y)$ is implicitly recoverable since $Y$ is fully observed.   $P(Y|I)$ may be recovered as shown below:
	\begin{align*}
	P(Y|I) & = P(Y|I,r'_I)\mbox{ (using $ Y \compos R_I|I$)} \nonumber \\
	&= P(Y^*=Y|I^*=I,,r'_I) \mbox{ (using equation \ref{eq:miss1})
	}
	\end{align*}
	
	Expressing $P(Y)=\sum_y P(Y|I)P(I)$ in matrix form, we get:\\
	\[ \left( \begin{array}{c}
	P(y') \\
	P(y)
	\end{array} \right)=
	\left( \begin{array}{cc}
	P(y'|i') & P(y'|i)\\
	P(y|i') & P(y|i)
	\end{array} \right)
	\left( \begin{array}{c}
	P(i') \\
	P(i)
	\end{array} \right)
	\]
	Assuming that the square matrix on R.H.S is invertible, $P(I)$ can be estimated as: \[\left( \begin{array}{cc}
	P(y'|i') & P(y'|i)\\
	P(y|i') & P(y|i)
	\end{array} \right)^{-1} \left( \begin{array}{c}
	P(y') \\
	P(y)
	\end{array} \right) \]
	Having recovered $P(I)$, the query $P(I,Y)$ may be recovered as $P(Y|I)P(I)$. 
	\label{ex:matrix}
\end{example}
General procedures for handling non-recoverable cases using both data and graph is discussed in \cite{mohan18handling}. The preceding recoverability procedure was inspired by similar results in causal inference \citep{pearl:09-r350, kuroki2014measurement}. In contrast to \cite{pearl:09-r350} that relied on external studies to compute causal effect in the presence of an unmeasured confounder, \cite{kuroki2014measurement} showed how the same could be effected without external studies. In missing data settings we have access to partial information that allows us to compute conditional distributions. This allows us to adapt the procedure in  \cite{pearl:09-r350} to establish recoverability. Yet another way of handling  these problems is based on double sampling wherein after the initial data collection a a random sample of non-respondents are tracked and their outcomes ascertained \citep{holmes2018estimated,zhang2016nonrespondent}.

\subsection{Recovering Causal Effects}
\label{sec:causal}

\begin{figure}
\begin{minipage}{.75\linewidth}
  \centering
  \includegraphics[scale=0.8]{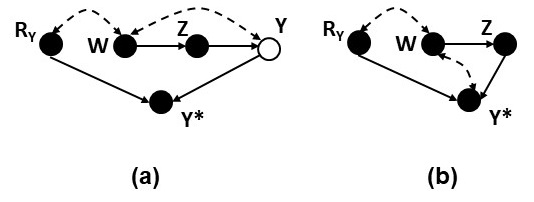}
  \label{fig:test1}
\end{minipage}%
\begin{minipage}{.25\linewidth}
  \centering
  \includegraphics[scale=1.05]{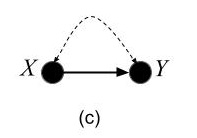}
  \label{fig:test2}
\end{minipage}
\caption{m-graphs in which (a) $P(y|do(z))$ is recoverable (b) $Y$ is treated as a latent variable and not explicitly portrayed. (c) bow-arc model in which causal effect of $X$ on $Y$ is non-identifiable.  }
	\label{fig:example}
\end{figure}

We assume the reader is familiar with the basic notions of
''causal queries", ''causal effect" and ''identifiability" as described in \cite{pearl} (chapter 3) and \cite{pearl:09-r350}.
Given a causal query and a causal graph with no missingness,
we can always determine whether or not the query is identifiable
using the \textit{complete} algorithm
in \cite{ilyabestpaper} or \cite{ huang2006identifiability} which
outputs an estimand whenever identifiability holds.
In the presence of missingness, a necessary condition for
recoverability of a causal query is its identifiability in
the substantive model i.e. the subgraph comprising of
$V_o,$ $V_m$ and $U$. In other words, a query which is not
identifiable in this model will not be recoverable under
missingness. A canonical example of such case is the bow-arc
graph (figure \ref{fig:example} (c)) for which the query $P(Y|do(X=x))$ is known to be
non-identifiable (\cite{pearl}) 
In the remainder of this
subsection  we will assume that queries of interest
are identifiable in the substantive model, and our task is
to determine whether or not they are recoverable from
the m-graph.  Clearly, identifiability entails the
derivation of an estimand, a sufficient condition for
recoverability is that the estimand in question be
recoverable from the m-graph.

\begin{figure}[h]
	\centering
	\includegraphics[scale=0.6]{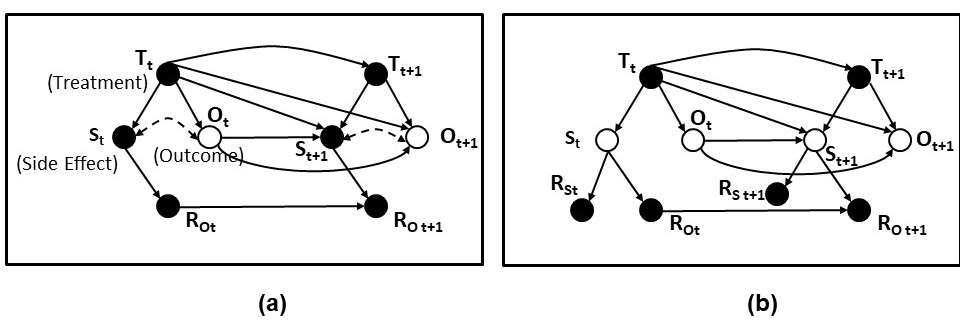}
	\caption{m-graphs depicting the problem of attrition. (a) MAR (b) MNAR
	}
	\label{fig:attrition}
\end{figure}

\begin{example} Consider the m-graph in
	in figure \ref{fig:attrition} (a), where it is required to
	recover the causal effect of two sequential treatments,
	$T_t$ and $T_{t+1}$ on outcome $O_{t+1}$, namely \\
	$P(O_{t+1}|do(T_t,T_{t+1})$.
	This graph models a longitudinal study with attrition, where
	the $R$ variables represent subjects dropping out of the study
	due to side-effects $S_t$ and $S_{t+1}$ caused by the corresponding
	treatments (a practical problem discussed in \cite{breskin2018practical,cinelli:pea18-r479}). The bi-directed
	arrows represent unmeasured
	health status indicating that participants with poor health are
	both more likely to experience side effects and incur
	unfavorable outcomes. Leveraging the exogeneity of
	the two treatments (rule 2 of do-calculus), we can remove
	the do-operator from the query expression, and obtain the
	identified estimand $P(O_{t+1}|do(T_t,T_{t+1})=P(O_{t+1}|T_t,T_{t+1})$.
	Since the parents of the $R$ variables are fully observed,
	the problem  belongs to the MAR category, in which
	the joint distribution is recoverable (using corollary \ref{cor:mar}).
	Therefore $P(O_{t+1}|T_t,T_{t+1})$ and hence our causal
	effect is also recoverable, and is given by:  $\sum_{S_t,S_{t+1}} P(O_{t+1}|T_t,T_{t+1},S_t,S_{t+1},R_{O_{t+1}}=0) P(S_t,S_{t+1}|T_t,T_{t+1})$.
\end{example}

Figure \ref{fig:attrition}(b) represents a more intricate variant of the
attrition problem, where the side effects themselves
are partially observed and, worse yet, they cause their
own missingness. Remarkably, the query is still recoverable,
using Theorem 1 and the fact that, (i) $O_{t+1}$ is
d-separated from both $R_{O_{t+1}}$ and $R_{O_t}$ given $(T_t, T_{t+1}, O_t)$, and (ii) $O_t$ 
is d-separated from  $R_{O_{t}}$ given $(T_{t}, T_{t+1})$. 
The resulting estimand is: $\sum_{O_t} P(O_{t+1}|T_t,T_{t+1},O_t,R_{O_t}=0,R_{O_{t+1}}=0) P(O_t|R_{O_t}=0,T_t,T_{t+1})$.

Figure \ref{fig:example}(a) portrays another example of identifiable query, but
in this case, the recoverability of the identified
estimand is not obvious; constraint-based analysis (\ref{sec:complex})
is needed to establish its recoverability.

\begin{example}
Examine the m-graph in figure \ref{fig:example}(a). Suppose we are interested in the causal effect of
Z (treatment) on outcome Y (death) where
treatments are conditioned on (observed) X-rays report
(W). Suppose that some unobserved factors
(say quality of hospital equipment and staff) affect both
attrition ($R_y$) and accuracy of test reports (W).
In this setup the causal-effect query $P(y|do(z))$ is
identifiable (by adjusting for W) through the estimand:
\begin{align}P(y|do(z))&= \sum_w P(y|z,w)P(w) \label{eq:v} \end{align}
However, the factor $P(y|z,w)$ is not recoverable (by theorem
\ref{thm:nonrecov}), and one might be tempted to conclude that
the causal effect is non-recoverable. We shall now
show that it is nevertheless recoverable in three steps.

\paragraph{Recovering $P(y|do(z)$ given the m-graph in figure
	\ref{fig:example}(a) }
The first step is to transform the query (using the rules of
do-calculus) into an equivalent expression such that no partially
observed variables resides outside the do-operator.
\begin{align}
P(y|do(z)) & =  P(y|do(z),R_y=0) \mbox{ (follows from rule 1 of
	do-calculus)} \nonumber \\
&=P(y^*|do(z),R_y=0) \mbox{ (using eq \ref{eq:miss1})}\label{eq:verma1}
\end{align}
The second step is to simplify the m-graph by removing
superfluous variables, still retaining all relevant functional
relationships. In our example $Y$ is irrelevant once we treating
$Y^*$ as an outcome. The reduced m-graph is shown in
figure \ref{fig:example}(b). The third step is
to apply the do-calculus (\cite{pearl})
	to the reduced graph (\ref{fig:example}(b)), and identify the modified query $P(y^*|do(z),R_y=0)$.
 \begin{align}
	P(y^*|do(z),R_y=0) &=  \sum_w P(y^*|do(z),w, R_y=0) P(w|do(z),R_y=0) \label{eq:a}  \\
	P(y^*|do(z),w, R_y=0) & = P(y^*|z,w,R_y=0)\label{eq:b}  \mbox{ (by Rule-2 of do-calculus) } \\
	P(w|do(z),R_y=0) & = P(w|R_y=0)\label{eq:c} \mbox{ (by Rule-3 of do-calculus) } ) \\	
	\intertext{Substituting (\ref{eq:b}) and (\ref{eq:c}) in (\ref{eq:a}) the causal effect becomes}
		P(y|do(z)) &= \sum_w P(y^*|z,w,R_y=0)P(w|R_y=0)\label{eq:v2}
\end{align}		
		which permits us to estimate our query from complete cases
		only. While in this case we were able to recover the causal effect using
		one pass over the three steps, in more complex cases we might need
		to repeatedly apply these steps in order to recover the query.
\end{example}

\begin{figure}[h]
	\centering
	\includegraphics[scale=1.5]{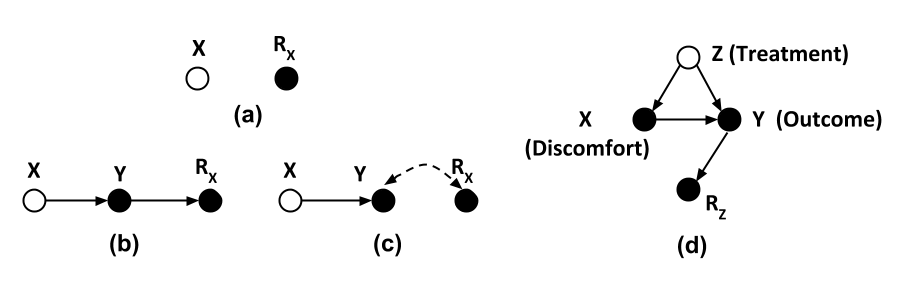}
	\caption{(a) m-graph with an untestable claim: $Z \compos R_z|X,Y$, (b) \& (c) Two statistically indistinguishable models, (d) m-graph depicting MCAR. }
	\label{fig:testability}
\end{figure}
\section{Testability Under Missingness}
\label{sec:testability}
In this section we seek ways to 
detect mis-specifications of the missingness model. While discussing  testability, one must note a phenomenon that recurs in missing data analysis: \emph{Not all that looks testable is testable.} Specifically, although every d-separation in the graph implies conditional independence in the recovered distribution, some of those independencies
are imposed by construction, in order to satisfy
the model's claims, and these do not provide means of refuting
the model. We exemplify this peculiarity below. 

\begin{example}Consider
the m-graph in figure \ref{fig:testability}(a). It is evident 
that the problem is MCAR (definition in section \ref{sec:mar}). Hence $P(X,R_x)$ is recoverable. The only conditional independence embodied in the graph is $X \compos R_x$. At first glance it might seem as if $X \compos R_x$ is testable since  we can go to the recovered
distribution and check whether it satisfies this
conditional independence. 
However, $X \compos R_x$ will always be satisfied
in the recovered distribution, because it was recovered
 so as to satisfy $X \compos R_x$. This can be shown explicitly as follows:
 
\begin{align*}
P(X,R_x) &= P(X|R_x)P(R_x)\\
&=P(X|R_x=0) P(R_x)\mbox{ (Using $X \compos R_x$)}\\
&=P(X^*|R_x=0) P(R_x) \mbox{( Using Equation \ref{eq:miss1})}
\intertext{ Likewise, }
P(X)P(R_x) &= P(X^*|R_x=0)P(R_x)
\end{align*} 

Therefore, the claim, $X \compos R_x$, cannot
be refuted by any recovered distribution, regardless of
what process actually generated the data.
In other words, any data
whatsoever with $X$ partially observed can be
made compatible with the model postulated.
\end{example}
The following theorem characterizes a more general class of untestable claims. 
\begin{thm}[\cite{mohan2013test}] Let $\{Z,X\} \subseteq V_m$ and $W \subseteq V_o$. 
	Conditional independencies of the form $X \compos R_x|Z, W,R_z$ are untestable. \label{thm:untestable}
\end{thm}
The preceding example demonstrates this theorem as a special case, with $Z=W=R_z=\emptyset$. 
The next section provides criteria for testable claims.	
\subsection{Graphical Criteria for Testability} \label{sec:syntax}
The criterion for detecting testable implications reads as follows:\textit{ A d-separation condition displayed in the graph is testable if the R variables associated with all the
partially observed variables in it are either present in the separating set or can be added to the separating set without
spoiling the separation. }
The following theorem formally states this criterion using three syntactic rules (\cite{mohan2013test}). 
\begin{thm}
A sufficient condition for an m-graph to be testable is that it encodes one of the following types of independences: 
\begin{align}
 & X \compos Y | Z, R_x, R_y, R_z \label{syntax:1}\\
 & X \compos R_y | Z, R_x, R_z \label{syntax:2}\\
 & R_x \compos R_y | Z, R_z \label{syntax:3}
\end{align}
\label{thm:test:sufficient}
\end{thm}
In words, any d-separation that can be expressed in the format stated above is  testable.
It is understood that, if $X$ or $Y$ or $Z$ are fully
observed, the corresponding $R$ variables may be removed from
the conditioning set. Clearly, any conditional independence
comprised exclusively of fully observed variables is testable. To search for such refutable claims, one needs to only examine the missing edges in the graph and check whether any of its associated set of separating sets satisfy the syntatctic format above.

To illustrate the power of the criterion we present the following example. 
\begin{example}
	Examine the m-graph in figure \ref{fig:testability} (d). The missing edges between $Z$ and $R_z$, and $X$ and $R_z$ correspond to the conditional  independencies: $Z \compos R_z|(X,Y)$ and $X \compos R_z|Y$, respectively. The former is untestable (following theorem \ref{thm:untestable}) while the latter is testable, since it complies
	with (\ref{syntax:2}) in theorem \ref{thm:test:sufficient}.
\end{example} 
\subsubsection{Tests Corresponding to the Independence Statements in Theorem \ref{thm:test:sufficient} }
A testable claim needs to be expressed in terms of proxy variables before it can be operationalized. For example, a specific instance of the claim
  $X \compos Y | Z, R_x, R_y, R_z$,  
 when $R_x=0, R_y=0,R_z=0$  gives $X \compos Y|Z, R_x=0, R_y=0, R_z=0$. On rewriting this claim as an equation and applying equation \ref{eq:miss1} we get,
	\begin{align*}
	P(X^*|Z^*, R_x=0, R_y=0, R_z=0)&=P(X^*|Y^*,Z^*, R_x=0, R_y=0, R_z=0)
	\end{align*}
	This equation  exclusively comprises of observed quantities and can be directly tested given the input distribution: $P(X^*,Y^*,Z^*,R_x,R_y,R_z)$. Finite sample techniques for testing conditional independencies
	are cited in the next section. In a similar manner we can devise tests for the remaining two statements in theorem \ref{thm:test:sufficient}.
	
The tests corresponding to the three independence statements in theorem \ref{thm:test:sufficient} are:
\begin{itemize}
	\item $P(X^*|Z^*, R_x=0, R_y=0, R_z=0)=P(X^*|Y^*,Z^*, R_x=0, R_y=0, R_z=0)$,
	\item $P(X^*|Z^*,R_x=0,R_z=0)=P(X^*|R_y,Z^*,R_x=0,R_z=0)$ 
	\item  $P(R_x|Z^*,R_z=0)=P(R_x|R_y,Z^*,R_z=0)$
\end{itemize}
The next section specializes these results to the classes of MAR and MCAR problems which have been given some attention in the existing literature. 
\subsection{Testability of MCAR and MAR}\label{sec:mar}
A chi square based test for MCAR was proposed by \cite{little1988test} in which a high value falsified MCAR\citep{Rubin_missing}. Rubin-MAR  is known to be untestable \citep{impute}. 
\cite{potthoff2006can} defined MAR at the variable-level (identical to that in section \ref{sec:mechanism}) and  showed that it can be tested.
Theorem \ref{thm:mar}, given below presents stronger conditions under which a given MAR model is testable (\cite{mohan2013test}). Moreover, it provides diagnostic insight in case the test is violated. We further note that these conditional independence tests may be implemented in practice using different techniques such as G-test, chi square test, testing for zero partial correlations or by tests such as those described in \cite{szekely2007measuring, gretton2012kernel, sriperumbudur2010hilbert}. 
\begin{thm}[MAR is Testable]\label{thm:mar}
	Given that $|V_m|>0$, $V_m \compos R | V_o$ is testable if and only if $|V_m|>1$ i.e. $|V_m|$ is not a singleton set.
\end{thm}
In words, given  a dataset with two or more partially observed variables, it is always possible to test whether MAR holds.  We exemplify such tests below.

\begin{example}[Tests for MAR] Given a dataset where $V_m=\{A,B\}$ and $V_o=\{C\}$, the MAR condition states that $(A,B) \compos (R_A,R_B)|C$. This statement implies the following two statements which match syntactic criteria in \ref{syntax:2} and hence are testable. 
		\begin{enumerate}
			\item $A \compos R_B|C,R_A$ 
			\item $B \compos R_A|C,R_B$
		\end{enumerate}
			The testable implication corresponding to (1) and (2) above are the following:
			\begin{align*}
			    			 P(A^*,R_B|C,R_A=0)&=P(A^*|C,R_A=0)P(R_B|C,R_A=0) \\
			    			 P(B^*,R_A|C,R_B=0)&=P(B^*|C,R_B=0)P(R_A|C,R_B=0)
			\end{align*}
\end{example} 
While refutation of these tests immediately imply that the data are not MAR,   we can never \textit{verify} the MAR condition. However if MAR  is refuted,  it is possible to pinpoint and locate the source of error in the model.  For instance, if claim (1) is refuted then one should consider adding an edge between $A$ and $R_B$.
\begin{remark} A recent paper by I Bojinov, N Pillai and D Rubin \citep{bojinov2017diagnosing}  has adopted some of the aforementioned
tests for MAR models, and demonstrated their use on simulated data.
Their paper is a testament to the
significance and applicability of our results
(specifically, section 3.1 and 6 in [36] to real world problems.
\end{remark}

\begin{cor}[MCAR is Testable]\label{cor:mcar}
	Given that $|V_m|>0$, $(V_m, V_O) \compos R | V_o$ is testable if and only if $|V_m| + |V_O| \ge 2$.
\end{cor}

\begin{example}[Tests for MCAR] Given a dataset where $V_m=\{A,B\}$ and $V_o=\{C\}$, the MCAR condition states that $(A,B,C) \compos (R_A,R_B)$. This statement implies the following statements which match syntactic criteria in \ref{syntax:2} and \ref{syntax:1} and hence are testable. 
		\begin{enumerate}
			\item $A \compos R_B|R_A$ 
			\item $B \compos R_A|R_B$
			\item $C \compos R_A$
		\end{enumerate}
			The testable implication corresponding to (1) and (2) above are the following:
			\begin{align*}
			    			 P(A^*,R_B|C,R_A=0)&=P(A^*|C,R_A=0)P(R_B|C,R_A=0) \\
			    			 P(B^*,R_A|C,R_B=0)&=P(B^*|C,R_B=0)P(R_A|C,R_B=0)\\
			    			 P(C,R_A)&=P(C)P(R_A)
			\end{align*}
\end{example}
\subsection{On the Causal Nature of the Missing Data Problem}
Examine the m-graphs in Figure \ref{fig:testability}(b) and (c).  $X \compos R_x|Y$ and $X \compos R_x$ are the conditional independence statements embodied in models \ref{fig:testability}(b) and (c), respectively. Neither of these statements are testable. Therefore they are statistically indistinguishable. However, notice that $P(XY)$ is recoverable in figure  \ref{fig:testability}(b) but not in  figure \ref{fig:testability}(c) implying that,
\begin{itemize}
	\item No universal algorithm exists that can decide if a query is recoverable or not without looking at the model. 
\end{itemize}
Further notice that $P(X)$ is recoverable in both models albeit using two different methods. 	In model \ref{fig:testability}(b) we have $P(X)=\sum_YP(X^*|Y,R_x=0)P(y)$ and in model \ref{fig:testability}(c) we have
	$P(X)=P(X^*|R_x=0)$. This leads to the conclusion that, 
	\begin{itemize}
		\item 	No universal algorithm exists that can produce a consistent estimate whenever such exists.	 
	\end{itemize}
The impossibility of determining from statistical assumptions alone, (i) whether a query is recoverable and (ii) how the query is to be recovered, if it is recoverable, attests to the causal nature of the missing data problem. Although \cite{Rubin_missing} alludes to the causal aspect of this problem, subsequent research has treated missing data mostly as a statistical problem. A closer examination of the testability and recovery conditions
shows however that a more appropriate perspective would be to treat
missing data as a causal inference problem.

\section{Conclusions}
All methods of missing data analysis rely on
assumptions regarding the reasons for missingness.
Casting these assumptions in a graphical model, permits
researchers to benefit from the inherent
transparency of such models as well as their ability to
explicate the statistical implication of the underlying
assumptions in terms of conditional independence relations
among observed and partially observed variables.
We have shown that these features of graphical models
can be harnessed to study unchartered territories of
missing data research.  In particular, we charted the
estimability of statistical and causal
parameters in broad classes of MNAR problems, and the
testability of the model assumptions under missingness
conditions.

An important feature of our analysis is its query
dependence. In other words, while certain properties of
the underlying distribution may be deemed unrecoverable,
others can be proven to be recoverable, and by smart
estimation algorithms.

We should emphasize that all our results assume
non parametric models. In other words, no assumptions
are needed about the functional or distributional nature
of the relationships involved.

In light of our findings we question the benefits of the traditional taxonomy that
classifies missingness problems into MCAR, MAR and MNAR. To decide if a problem falls into any
of these categories a user must have a model of the causes of missingness and once this model is articulated 
the criteria we have derived for recoverability and testability can be readily applied. Hence we see no need to 
refine and elaborate conditions for MAR. 

The testability criteria derived in this paper
can be used not only to rule out misspecified models
but also to locate specific mis-specifications for the purpose
of model updating and re-specification. More importantly, we
have shown that it is possible to determine if and how a
target quantity is recoverable, even in models where
missingness is not ignorable.
Finally, knowing which sub-structures in the graph prevent
recoverability can guide data collection procedures by identifying
auxiliary variables that need to be measured to ensure
recovery, or problematic variables that may compromise
recovery if measured imprecisely.

\bibliographystyle{chicago}
\bibliography{final}
\newpage
\section{Appendix}
\subsection{Estimation when the Data May not be Missing at Random. (\cite{Rubin2014}, page-22)}
\label{app:rubin}
Essentially all the literature on multivariate incomplete data assumes that the data
are MAR, and much of it also assumes that the data are MCAR. Chapter 15 deals
explicitly with the case when the data are not MAR, and models are needed for the
missing-data mechanism. Since it is rarely feasible to estimate the mechanism with
any degree of confidence, the main thrust of these methods is to conduct sensitivity
analyses to assess the effect of alternative assumptions about the missing-data
mechanism.
\subsection{A Complex Example of Recoverability}
\label{sec:complex}
We use $R=0$ as a shorthand for the event where all variables are observed i.e. $R_{V_m}=0$.
\begin{example} Given the m-graph in figure \ref{fig:inducing} (c), we will now recover the joint distribution.
	\begin{align}
	P(W,X,Y,Z)&=P(W,X,Y,Z)\frac{P(W,X,Y,Z,R=0)}{P(W,X,Y,Z,R=0)}=\frac{P(W,X,Y,Z, R=0)}{P(R=0|W,X,Y,Z)} \nonumber \\
	\intertext{Factorization of the denominator based on topological ordering of $R$ variables yields,}
	P(W,X,Y,Z)&=\frac{P(W,X,Y,Z, R=0)}{P(R_y=0|W,X,Y,Z,R_x=0,R_w=0,R_z=0)P(R_x=0|W,X,Y,Z,R_w=0,R_z=0)}\nonumber \\
	&\hspace*{5cm}\frac{1}{P(R_w=0|W,X,Y,Z,R_z=0)P(R_z=0|W,X,Y,Z)} \nonumber\\
	\intertext{On simplifying each factor of the form: $P(R_a=0|B)$, by removing from it all $B_1 \in B$ such that $R_a \compos B_1|B-B_1$, we get:}
	P(W,X,Y,Z)&=\frac{P(W,X,Y,Z, R=0)}{P(R_z=0)P(R_w=0|Z)P(R_y=0|X,W,R_x=0)P(R_x=0|Y,W)} \label{ex:eq:gmf}
	\end{align}
	$P(WXYZ)$ is recoverable if all factors in the preceding equation is recoverable. Examining each factor one by one we get:
	\begin{itemize}
		\item $P(W,X,Y,Z, R=0)$: Recoverable as $P(W^*,X^*,Y^*,Z^*, R=0)$ using equation \ref{eq:miss1}.
		\item $P(R_z=0)$: Directly estimable from the observed-data distribution.
		\item $P(R_w=0|Z)$: Recoverable as $P(R_w=0|Z^*,R_z=0)$, using $R_w \compos R_z|Z$ and equation \ref{eq:miss1}.
		\item $P(R_y=0|X,W,R_x=0)$: Recoverable as $P(R_y=0|X^*,W^*,R_x=0,R_w=0)$, using $R_y \compos R_w|X,W,R_x$ and equation \ref{eq:miss1}.
		\item $P(R_x=0|Y,W)$: The procedure for recovering $P(R_x=0|Y,W)$ is rather involved and requires converting the probabilistic sub-query to a causal one as detailed below.
	\end{itemize}  
	\begin{align*}
	P(R_x=0|Y,W=w) &= P(R_x=0|Y,do(W=w)) \mbox{(Rule-2 of do calculus)}\\
	&=\frac{P(R_x=0|Y,R_y=0,do(w))}{P(R_x=0|Y,R_y=0,do(w))}P(R_x=0|Y,do(W=w))\\
	&= P(R_x=0|Y,R_y=0,do(w))\frac{P(R_y=0|Y,do(w))}{P(R_y=0|Y,do(w),R_x=0)} \numberthis \label{eq:causal}
	\end{align*}
	To prove recoverability of $P(R_x=0|Y,W=w)$, we have to show that all factors in  equation \ref{eq:causal} are recoverable. 
	\paragraph{Recovering $\mathbf{P(R_y=0|Y,do(w),R_x=0):}$} Observe that $P(R_y=0|Y,do(w),R_x=0) = P(R_y=0|do(w),R_x=0)$ by Rule-1 of do calculus. To recover $P(R_y=0|do(w),R_x=0)$ it is sufficient to show that $P(X^*,Y^*,R_x,R_y,Z|do(w))$ is recoverable in $G'$, the latent structure corresponding to $G$ in which $X$ and $Y$ are treated as latent variables.
	\begin{align*}
	P(X^*,Y^*,R_x,R_y,Z|do(w)) &= P(X^*,Y^*,R_x,R_y|Z,do(w))P(Z|do(w))\\
	&= P(X^*,Y^*,R_x,R_y|Z,w)P(Z|do(w)) \mbox{ (Rule-2 of do-calculus)}\\
	&= P(X^*,Y^*,R_x,R_y|Z,w)P(Z) \mbox{ (Rule-3 of do-calculus)}\\
	\intertext{Using $(X^*,Y^*,R_x,R_y) \compos (R_z,R_w )|(Z,W)$,  equation \ref{eq:miss1} and $Z\compos R_z$ we show that the causal effect is recoverable as:}
	P(X^*,Y^*,R_x,R_y,Z|do(w)) &= P(X^*,Y^*,R_x,R_y|Z^*,w^*,R_w=0,R_z=0)P(Z^*|R_z=0) \numberthis \label{eq:final1}
	\end{align*}
	\paragraph{Recovering $\mathbf{P(R_x=0|Y,do(w),R_y=0):}$} Using equation \ref{eq:miss1}, we can rewrite $P(R_x=0|Y,do(w),R_y=0)$ as $P(R_x=0|Y^*,do(w),R_y=0)$. Its recoverability follows from equation \ref{eq:final1}.
	\paragraph{Recovering $\mathbf{P(R_y=0|Y,do(w)):}$}
		\begin{align*}
	P(R_y=0|Y,do(w)) &= \frac{P(R_y=0,Y|do(w))}{\sum_{R_x}P(R_y=0,Y,R_x|do(w))+ P(R_y=1,Y,R_x|do(w))}\\
	&= \frac{P(R_y=0,Y^*|do(w))}{\sum_{R_x}P(R_y=0,Y^*,R_x|do(w))+ P(R_y=1,Y,R_x|do(w))} \mbox{(using eq \ref{eq:miss1})}\\
	\intertext{ $P(R_y=0,Y^*|do(w))$ and $P(R_y=0,Y^*,R_x|do(w))$ are recoverable from equation \ref{eq:final1}. We will now show that $P(R_y=1,Y^*,R_x|do(w))$ is recoverable as well. }
	P(R_y=1,Y,R_x|do(w)) &= \frac{P(R_y=0,Y,R_x|do(w))}{P(R_y=0|R_x,Y|do(w))}-P(R_y=0,R_x,Y|do(w))\\
	\intertext{Using equation \ref{eq:miss1} and Rule-1 of do-calculus we get,}
	&= \frac{P(R_y=0,Y^*,R_x|do(w))}{P(R_y=0|R_x,do(w))}-P(R_y=0,R_x,Y^*|do(w))
	\intertext{Each factor in the preceding equation is estimable from equation \ref{eq:final1}. Hence $P(R_y=1,Y,R_x,do(w))$ and therefore, $P(R_y=0|Y,do(w))$ is recoverable.}
	\end{align*}
	Since all factors in equation \ref{eq:causal} are recoverable, joint distribution is recoverable. 
\end{example}
\end{document}